\documentstyle[aps,twocolumn,epsf,rotate]{revtex}
\begin{document}
\draft
\twocolumn[\hsize\textwidth\columnwidth\hsize\csname
@twocolumnfalse\endcsname

\title{
Crossover from BCS superconductivity to BEC of pairs: The role of the
lifetime of the pairs.
}

\author{M. Letz}
\address{Institut f\"ur Physik,
Johannes-Gutenberg Universit\"at, 55099 Mainz, Germany}

\date{\today}

\maketitle

\begin{abstract}

The understanding of an electron gas with short coherence length pairs
formed by an attractive interaction is believed to be one of the major
keys to our theoretical knowledge of the high-T$_c$-superconductors. Mainly
the deviations of the cuprates from usual metallic Fermi liquid
behaviour already in the normal  state like e.g. a linear resistivity
or the observation of a 
pseudo gap can result from electron--electron correlations. 

We therefore investigate the negative U Hubbard model in two
dimensions at low densities using the T-matrix approximation. In the
non selfconsistent formulation of the theory the system always shows
an instability 
towards Bose condensation of pairs into an infinite lifetime
two--particle bound state. If the calculations are performed
selfconsistently pair--pair scattering is included which causes the
pairs to have finite lifetime. The physics of these finite lifetime
pairs is discussed.


\end{abstract}
]
\narrowtext 

\section{Introduction:}
\label{sec:intro}
More than a decade ago the high-T$_c$ superconductors have been
discovered. Even if many features of these materials like e.g. the
mainly two dimensional transport in the hole doped CuO$_2$ planes are understood
the microscopic mechanisms and models which lead to a wider
understanding still need to be investigated.\\[0.1cm]
Already in the normal state above T$_c$ there are two important issues
which need to be explained. The first one (I) is the linear
resistivity. In a normal metal the low-temperature dependence of the
resistivity is governed by either impurity scattering which gives a
constant term or by a Bloch-phonon contribution ($\sim$ T$^5$) or in
very clean Fermi liquids by the Landau electron--electron scattering of
electrons at the Fermi surface ($\sim$ T$^2$). Only at high
temperatures above the Debye temperature which is of the order of
300-400 K normal metals show a linear resistivity due to phonon
contributions.\\
In the high-T$_c$'s however the optimal doped samples show a linear
resistivity over the full temperature range, already well below the
Debye temperature \cite{linres}.\\
The second feature (II) in the normal state which needs to be explained
is the occurrence of a pseudo gap. It has been found in many
experiments (first in NMR \cite{warren89}) and also with other methods
e.g. tunneling experiments \cite{stm_pseu}. This pseudo gap is found
in the underdoped materials were already at temperatures above the
superconducting transition temperature T$_c$ a gap opens at the 
Fermi-surface.\\[0.1cm]
Also in the superconducting state there is besides the high transition
temperature another special feature of the cuprates. This is the short
coherence length of the pairs. Usual weak coupling BCS
superconductors have coherence lengths $\xi$ of several hundred lattice
constants ($\sim$ 1000 $\AA$) which go together with high quasiparticle
densities. Therefore many Cooper pairs ($\sim$ 10$^6$) overlap and
the two essential conditions for superconductivity, pairing and
phase--coherence, occur simultaneously at the same temperature, the
mean--field (BCS) T$_c$.\\
In the high-T$_c$ superconductors however the coherence length is very
small ($\xi \approx $ 10 $\AA$)  of the order of only 3-4 lattice
constants. Further the quasiparticle densities are very low since the
materials are close to an antiferromagnetic insulator to metal
transition. Therefor the pairs barely overlap and might even be
treated as spatially well separated pairs. This is believed to be the
reason \cite{emeryhoust} why pairing (opening of a pseudo gap?) and
phase coherence (T$_c$) might occur at different temperatures.\\ 
All these features above motivate the study of a model system which
allows to investigate the transition between usual BCS
superconductivity to a situation were tightly bound pairs occure.
The simplest model system which can describe this is the Hubbard model
with an attractive interaction between the electrons.

\section{Model:}
\label{sec:model}
The negative U Hubbard model is: 
\begin{equation}
\label{eq:hub}
H = t \sum_{<i,j>} c^{\dagger}_i c_j + U \sum_i  n_i n_j
\end{equation}
were the kinetic energy is given by the transfer term $t$ and the
Hubbard attraction $U$ is chosen to be negative.\\
In the weak coupling case in three dimensions the Fourie transform of
eq. (\ref{eq:hub}) describes just the starting Hamiltonian for the BCS
theory.\\
In the zero density and strong coupling (atomic) limit ($n/U$ = 0) the
Hamiltonian is solved by local pairs of electrons whose energy is lowered by
the attractive energy U. Such pairs of electrons can be considered as
Bosons.\\
In the intermediate coupling regime ($|U| \approx$ bandwidth W) and
for low densities $n$ short coherence length pairs which barely
overlap can be described. This is the parameter regime we are
interested in in the current work.

\section{T-Matrix:}
\label{sec:t-matrix}
An approximation which allows to access the low density regime and
which also describes pair formation is the T-matrix approximation. It
is often also called ladder-approximation or Br\"uckner-Hartree-Fock
theory. In this theory the infinite sum over all scattering events
between two electrons enters in the vertex function $\tilde{\Gamma}$.
\begin{equation}
\label{eq:gamt}
\tilde{\Gamma}({\bf K}, i \Omega _n) = 
\frac{U}{1-U \; \chi({\bf K}, i \Omega _n)}
\end{equation}
With ${\bf K}$ being the total momentum of a pair and $ \Omega _n$
Bosonic Matsubara frequencies. The susceptibility 
$\chi({\bf K}, i \Omega _n)$ is given by the product of two
one-particle Green's functions $G({\bf k},i \omega _m)$.
\begin{equation}
\label{eq:chit}
\chi({\bf K}, i \Omega _n)
 = \frac{-1}{N \beta} \sum_{m,{\bf k}}  
G({\bf K}-{\bf k},i \Omega _n - i \omega _m) G({\bf k},i \omega _m) 
\end{equation}
Spin indices are left out since only singlet pairs can be formed with
an on-site attraction.

\subsection{Thouless-criterion:}
\label{sec:thoul}
The physics which is described with the T-matrix approximation depends
crucially on the appearance of a two particle bound state which is
given by a zero 
of the denominator of eq. (\ref{eq:gamt}) reaches the chemical 
potential first at ${\bf K }= {\bf
0}$ and is called the Thouless instability \cite{thoulesskrit}.
\begin{equation}
\label{eq:bs}
\chi({\bf K}={\bf 0}, i \Omega _n) = \frac{1}{U}
\end{equation}
Here we have to distinguish between two different cases:\\
In the first one (I) the chemical potential is in the one particle
continuum. One has a Fermi surface. This happens in the weak coupling
3D case when a BCS instability is formed at the Fermi surface. The
Thouless instability occurs when the system is cooled down
towards T$_c$. Eq. (\ref{eq:bs}) is (if $\chi=\chi^0$ is build up from
noninteracting one-particle Green's functions $G^0$) identical to the
equation which determines T$_c$ in the BCS theory.\\
In the second case (II) the chemical potential is below the one-particle
continuum. 
When considering $\chi^0$ this can always lead to a bound state at
${\bf K }= {\bf 0}$ in 1D and 2D. In 3D this condition only leads to a
bound state below the one-particle continuum if $|U| > U_{crit}$. Such
a bound state can be populated by pairs of electrons which obey Bose
statistics and therefore Bose condensation can occur. A Fermi
surface will be lost at low temperatures. The ${\bf K}$-dispersion of
such a bound state is shown in fig. \ref{fig:bs} (a).

\subsection{The selfenergy:}
\label{sec:self}
The vertex function can now be used to build up a full Green's
function and different degrees of selfconsistency are possible. In
this work we only want to focus on two ways to close the equations. 

\subsubsection{Non selfconsistent calculation} 
The
first one is the non-selfconsistent calculation (see fig. \ref{fig:G4NSCC}):
\begin{figure}
\unitlength1cm
\epsfxsize=1.8cm
\begin{picture}(7,3)
\put(-0.4,0.5){\rotate[r]{\epsffile{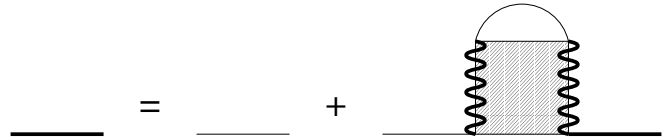}}}
\end{picture}
\caption{Diagram for the single-particle Green's function (solid line) in the 
non self-consistent approximation. The thin solid lines
represent the non-interacting Green's function, the thick lines fully
interacting Green's functions and the wavy lines
represent the interaction U.}
\label{fig:G4NSCC}
\end{figure}
\begin{equation}
\label{eq:greencons}
G({\bf k},i \omega _n) = \left ( G^0({\bf k},i \omega _n)^{-1} - 
\Sigma^0({\bf k},i
\omega _n) \right ) ^{-1}
\end{equation}
were the selfenergy $\Sigma^0({\bf k},i \omega _n)$ is build up from
the vertex function:
\begin{equation}
\Sigma^0({\bf k},i \omega _n) =  \frac{1}{N \beta} \sum_{m,{\bf q}}
\Gamma^0({\bf k}+{\bf q },i \omega _m + 
i \omega _n) G^0({\bf q},i \omega _m) \label{eq:sigscf} 
\end{equation} 
The system now consists of pairs of electrons which can only thermally
be excited and can at finite temperatures be described as a mixture of
Bosons (pairs) and Fermions. At zero temperature Bose condensation of
the pairs into the two-particle bound state always occurs. This has
been shown by Schmitt-Rinck et al. 
\cite{svr} were a version of the T-matrix theory was applied which was
not even conserving on a one-particle level.

\subsubsection{selfconsistent calculation}
\begin{figure}
\unitlength1cm
\epsfxsize=10.7cm
\begin{picture}(7,9.5)
\put(-3,0){\rotate[r]{\epsffile{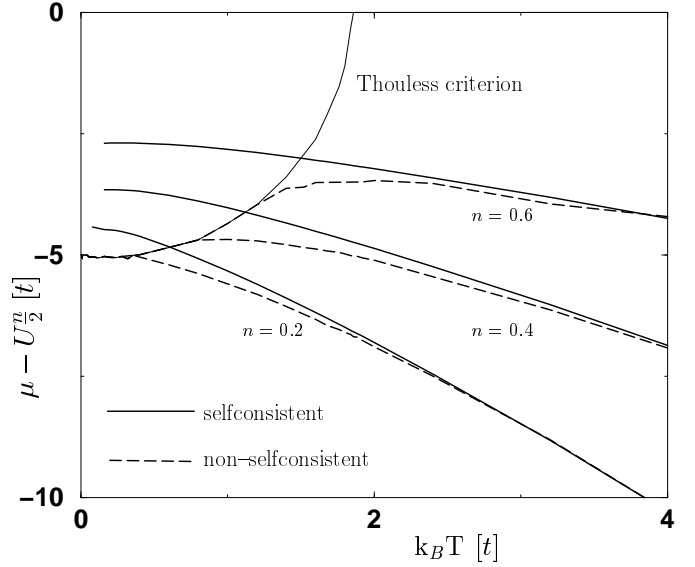}}}
\end{picture}
\caption{
The chemical potential is plotted as a function of temperature. The
calculation was done for the 
interaction  $U=-8t$ on a 12 by 12 lattice using 150 frequency points
along the real axis. The dashed lines are results from
non--selfconsistent calculations. At low temperatures they always show
Bose--condensation of pairs into the two--particle bound state below
the line given by the non--selfconsistent Thouless criterion. In the
selfconsistent calculation we always regain a Fermi--like surface, but
never reach a Thouless instability. For this plot the Hartree term was
left out.
}
\label{fig:flow}
\end{figure}

The second possibility we want to discuss here is the fully
selfconsistent version of the T-matrix. 
Other degrees of selfconsistency have been discussed e.g. by Janko et
al. \cite{janko98}.
Why we concentrate our discussion on the fully selfconsistent T-matrix
calculation will be investigated elsewhere \cite{frankm98}.
\begin{figure}
\unitlength1cm
\epsfxsize=1.8cm
\begin{picture}(7,3)
\put(-0.4,0.5){\rotate[r]{\epsffile{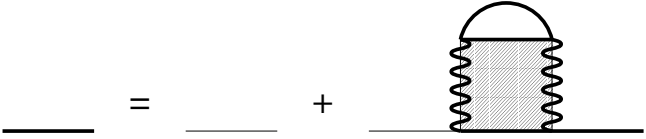}}}
\end{picture}
\caption{Diagram for the single-particle Green's function in the 
self-consistent, conserving approximation, with the same notation
as fig. 1.}
\label{fig:G4SCC}
\end{figure}
The full Green's function is now given by (see fig. \ref{fig:G4SCC}):
\begin{equation}
\label{eq:greenscf}
G({\bf k},i \omega _n) = \left ( G^0({\bf k},i \omega _n)^{-1} - 
\Sigma({\bf k},i
\omega _n) \right ) ^{-1}
\end{equation}
were the selfenergy $\Sigma({\bf k},i \omega _n)$ is build up from
the vertex function which is formed from a susceptibility being a
product of full one-particle Green's functions as well:
\begin{equation}
\Sigma({\bf k},i \omega _n) =  \frac{1}{N \beta} \sum_{m,{\bf q}}
\Gamma({\bf k}+{\bf q },i \omega _m + 
i \omega _n) G({\bf q},i \omega _m)  
\end{equation} 
In the zero density limit ($n/U=0$) eqs. (\ref{eq:greencons}) and
(\ref{eq:greenscf}) are identical. However at finite densities a
different physical scenario is been described by the selfconsistent
calculation. This results from the fact that pair--pair interactions
are now included in eq. (\ref{eq:greenscf}). In this way the pairs can
already at
zero temperature have finite lifetime, they therefore deviate
from Bose statistics and therefore the Bose condensation is strongly
hindered (see fig. \ref{fig:flow}).

\subsection{Calculation procedure}
   
In order to solve the system of selfconsistency equations we use a
spectral representation for all correlation functions. The vertex
function is turned into an analytic function 
$\Gamma({\bf K}, i \Omega_n)$
by subtracting the
Hubbard energy U which is identical to subtracting the Hartree term
from the self energy.  
\begin{equation}
\Gamma({\bf K}, i \Omega _n) =
\tilde{\Gamma}({\bf K}, i \Omega _n) - U = 
\frac{U^2 \chi({\bf K}, i \Omega _n)}{1-U \; \chi({\bf K}, i \Omega _n)}
\end{equation}
We approximate the functions $G({\bf k}, i \omega _n)$, $\chi({\bf K},
i \Omega _n)$, $\Gamma({\bf K}, i \Omega _n)$ and $\Sigma({\bf k}, i
\omega _n)$ by a series of delta functions along the real axis e.g.:
\begin{equation}
G({\bf k}, i \omega _n) = \sum_j \frac{a_j^{\bf k}}{i \omega_n -b_j}
\end{equation}
to obtain a selfconsistent solution we have to determine the coefficients 
$a_j^{\bf k}$ in an iterative way until selfconsistency is
achieved. The details of the method, especially how to choose the
frequency points $b_j$ are explained in appendix A.

\section{results}

\begin{figure}
\unitlength1cm
\epsfxsize=10.5cm
\begin{picture}(7,9.5)
\put(-2.5,0){\rotate[r]{\epsffile{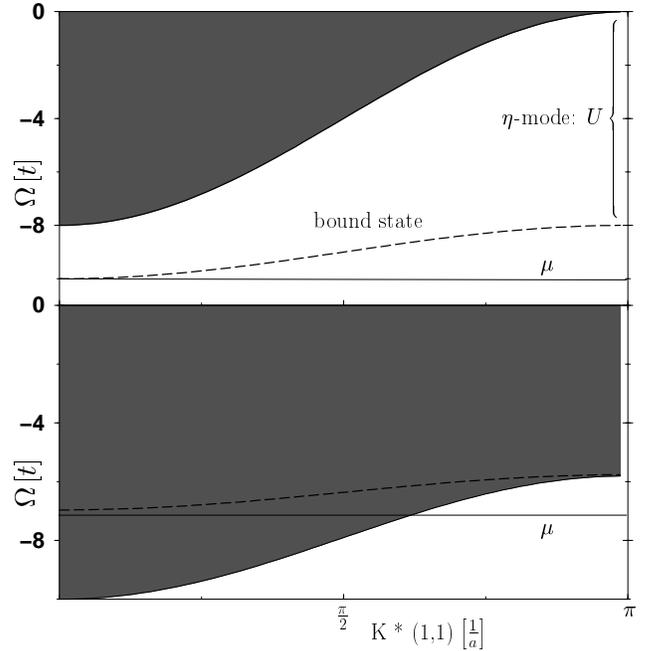}}}
\end{picture}
\caption{Schematic of the dispersion of the two-particle bound
state below the non-interacting continuum of the susceptibility 
$\chi(K,\Omega)$ (shaded region) as it arises in a non selfconsistent 
formulation (a).
At low enough temperatures the chemical potential ends always at the
${\bf K}=0$ two--particle bound state. Below (b) we have plotted the
same picture as it arises from a selfconsistent calculation done for
the density n=0.4, temperature k$_B$T=0.16[t] and interaction  $U=-8t$
on a 12x12 lattice. The two particle 
bound state is disappeared and only a strong weakly dispersive 
two--particle resonance directly above the chemical potential (dashed
line) is present.
}
\label{fig:bs}
\end{figure}

We have solved the equations on a 2D system, a finite lattice with
periodic boundary conditions. The results we present in the following
were obtained on a 12x12 lattice. The attractive interaction was
chosen to be equal to the bandwidth $U=-8t$.
At finite densities we find completely different physics if we compare
the selfconsistent and the non-selfconsistent calculations. In
fig. \ref{fig:flow} we have plotted the chemical potential as a
function of Temperature. Note that the Hartree term is still ignored. 
In the non--selfconsistent calculation the chemical potential always
ends up in the two--particle bound state, roughly
$-\frac{1}{2}\sqrt{U^2+\Delta^2}$ (with $\Delta$ the bandwidth) below
the middle of the unperturbed band. If we solve the T-matrix
selfconsistently we get the chemical potential back into the
one-particle continuum, we regain a Fermi--like surface but we never
reach a superconducting instability. Note that the Thouless criterion
in fig. \ref{fig:flow} refers only to the superconducting instability
of a non selfconsistent calculation.
We further get quasiparticle
peaks at ${\bf k}_F$ in the one--particle density of states whereas
the non--selfconsistent calculation shows (at low enough temperatures)
a gap at the chemical potential which is often interpreted as the
origin of the pseudo--gap \cite{janko98}. For a large temperature
range we further find a linear dependence of the quasiparticle
scattering rate with temperature \cite{linres98} which may explain
the linear resistivity.\\
In the current work we want to focus on two--particle properties. In
the non--selfconsistent version of the T-matrix a two--particle bound
state is present. At ${\bf K} = (\pi, \pi)$ it evolves into which is
called the
$\eta$-mode \cite{etamode96}. This is shown in fig. \ref{fig:bs}
(a). 
In order to answer the question how remnants of the two--particle
bound state show up in the selfconsistent calculation we have plotted
fig. \ref{fig:bs} (b). In this figure we show that the only remnant of
the bound state is a weakly dispersive, strongly lifetime broadened 
 resonance in the two--particle correlation function directly above
the chemical potential. It is remarkable that also when Boson--Fermion
models are discussed the Bosonic energy level is mostly put directly
above the chemical potential \cite{ranninger98}.
Also the $\eta$--mode only partially survives
the selfconsistent calculation. Even at ${\bf K} = (\pi, \pi)$ the
two--particle resonance is strongly lifetime broadened and lies now
less than $U$ below the middle of the unperturbed band (here zero).
At zero density however we regain the $\eta$--mode and the two
solutions for the selfconsistent and non--selfconsistent calculation
fall together. 
\begin{figure}
\unitlength1cm
\epsfxsize=10.5cm
\begin{picture}(7,8.5)
\put(-2.5,0){\rotate[r]{\epsffile{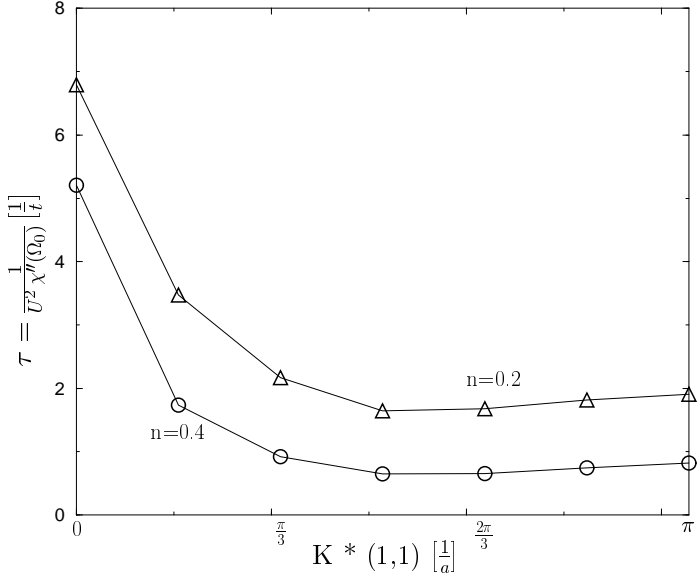}}}
\end{picture}
\caption{Lifetime of the pair for the selfconsistent calculation above
with the same 
parameters as in figs. 3 and 4. We have compared two different
densities. With decreasing density the lifetime of the pairs strongly
increases until it reaches at n=0 infinity were also the $\eta$-mode
will be recovered.
}
\label{fig:lifet}
\end{figure}
To quantify how strongly the pairs are
lifetime-broadened we have plotted in fig. \ref{fig:lifet} the
dispersion of the pair lifetime which is at a first approximation
given by:
\begin{equation}
\tau = \frac{1}{U^2 \; \chi''(\Omega_0)}
\end{equation}
were $\Omega_0$ is the frequency at which the two--particle resonance
occurs. The positions of the poles and the imaginary part of the
susceptibility at that points were extracted from our numerical results.
At ${\bf K} = (0,0)$ the pairs live longest. With decreasing particle
number the lifetime of the pairs increases until it reaches infinity
at n=0 were the selfconsistent and the non--selfconsistent solutions
fall together. 

\section{Conclusion}

We have shown in this work that a selfconsistent inclusion of
pair--pair scattering into the T-matrix calculation strongly alters
the physics which is described by this theory. Due to selfconsistency
the pairs can interact with each other which causes them to have
finite lifetime. These finite lifetime pairs can no longer Bose
condensate and therefore Fermi-liquid-like properties are regained. 
Although similar tendencies have been found earlier by other authors
\cite{fresard92,micnas95,haussmann93} we are able to work out many
details and to reach very small temperatures. This is due to the fact
that we use a semi-analytical method were all frequency integrations
are done analytically and only the k-space summation has to be done
numerically. Since we find the same results for finite lattices and
for the continuum in infinite dimensions \cite{letz98} we do not
believe that our results in general are sensitive to finite size effects.

\section{Acknowledgement}

I wish to thank F. Marsiglio and R.~J.~Gooding for many intense
discussions. I further 
acknowledge discussions with R.~Micnas, P.~Schuck and H.~C.~Ren.

\section*{appendix A: Selfconsistent procedure}

\label{appenda}
In order to solve the selfconsistent equations we apply a method
which enables us to work entirely along the real frequency axis. 
We do an Ansatz for the correlation functions (here e.g. $G({\bf k}, i \omega _n))$ of the
following form.
\begin{equation}
G({\bf k}, i \omega _n) = \sum_j^N \frac{a_j^{\bf k}}{i \omega_n -b_j}
\end{equation}
This means the spectral representation of $G({\bf k}, i \omega _n)$ is
approximated by a series of $\delta$ functions. In this work the frequencies
$b_j$ were kept fixed throughout the whole calculation (opposite to our
previous work \cite{letz98}). In order to access low
temperatures and work with a finite number of frequencies we sampled
the frequency points with a $\tanh$ function. In this way we make sure
that at the chemical potential two frequency points are always closer
to each other then k$_B$T.\\
On the example of the susceptibility we want to show how we can
calculate the spectral representation of a product function from the
one-particle Green's functions:
\begin{equation}
\label{eq:chi}
\chi({\bf K},i \Omega _n) = -\frac{1}{N \beta} \sum_{m,{\bf k}}  
G({\bf K}-{\bf k},i \Omega _n - i \omega _m) G({\bf k},i \omega _m) 
\end{equation}
When inserting the spectral representation for $G({\bf k}, i \omega _n)$ we get:
\begin{eqnarray}
\label{chispr}
\lefteqn{
\chi({\bf K}, i \Omega _n) = - \sum_{\bf k} \sum_{j,l}^{N} \frac{1}{\beta} \sum_m 
\frac{a_j^{{\bf K}-{\bf k}}}{i \omega _m - b_j} 
\frac{a_l^{{\bf k}}}{i \Omega _n - i \omega_m - b_l}} \nonumber \\
&=&
\sum_{\bf k} \sum_{j,l}^{N}  
\frac{a_j^{{\bf K}-{\bf k}} \; a_l^{\bf k}} {i \Omega _n - b_j - b_l} \left 
(
\frac{1}{1+e^{- \beta b_j}} - \frac{1}{1+e^{- \beta b_l}} \right )  
\nonumber \\ &=&
\frac{1}{2} \sum_{\bf k} \sum_{j,l}^{N} \frac{a_j^{{\bf K}-{\bf k}} \; a_l^{\bf k} }{i \Omega _n - b_j -
b_l} 
\nonumber \\ && \;\;\;\;\;\;\;\;\;\;\;\;\;\;\;\;
\left ( \tanh \left ( \frac{\beta b_j}{2} 
\right ) + \tanh \left ( \frac{\beta b_l}{2} 
\right ) \right )  
\end{eqnarray} 
We now have determined $\chi({\bf K}, i \Omega _n)$ on $N(N+1)/2$ frequency points. These
have to be sorted numerically and distributed onto the nearest
frequency points $b_j$. In order to calculate $\Gamma({\bf K}, i
\Omega _n)$ from $\chi({\bf K}, i \Omega _n)$ we
have to apply a numerical broadening which must again be smallest at
the chemical potential. In this way we get a complex function $\chi({\bf K}, i \Omega _n)$
from which we calculate the complex function $\Gamma({\bf K}, i \Omega _n)$. From the
imaginary part of $\Gamma({\bf K}, i \Omega _n)$ we calculate the amplitudes for the spectral
representation for $\Gamma({\bf K}, i \Omega _n)$. Finally we get:
\begin{equation}
\Gamma({\bf K}, i \Omega _n) = \sum_j^N \frac{g_j^{\bf K}}{i \Omega_n -b_j}
\end{equation}
A similar set of equations has to be solved for $\Sigma({\bf k}, i \omega _n)$ and $G({\bf k}, i \omega _n)$. In
this way the selfconsistency is closed. We defined our solution to be
selfconsistent if the mean--square deviation of all the amplitudes
$a_j^{\bf k}$ 
between two consecutive selfconsistency steps
was below a certain threshold.




\end{document}